\documentclass[twocolumn,epsfig]{revtex4}
\usepackage{epsfig,color}
\begin{document}
\title{$K^{+}$ and $K^{-}$ potentials in hadronic matter can be observed}

\author{Aman D. Sood$^1$ }
\email{amandsood@gmail.com}
\author{Ch. Hartnack$^1$ }
\author{J\"org Aichelin$^1$ }
\address{
$^1$SUBATECH,
Laboratoire de Physique Subatomique et des
Technologies Associ\'ees \\University of Nantes - IN2P3/CNRS - Ecole des Mines
de Nantes 
4 rue Alfred Kastler, F-44072 Nantes, Cedex 03, France}
\date{\today}

\maketitle

\section*{Introduction} 
One key question in the analysis of sub-threshold kaon production is how to obatin information on the properties of strange mesons in dense nuclear matter \cite{Aichelin:1986ss}. The principal problem for extracting  precise information on these properties  is, however, that almost all observables depend simultaneously not only on the $K^-$ potential but also on several other input quantities which are only vaguely known eg. life time of $\Delta$ and in-medium modification of the cross section. 
The situation were much better if experiment  provides an observable which depends on the $K$ potentials only 
and which is not spoiled by other little or unknown quantities. Here we aim show that the 
ratio of the $K^+$ and $K^-$  momentum spectra at small momentum in light systems can be such an observable.
\begin{figure}[!t] \centering
\vskip 0.5cm
\includegraphics[angle=0,width=6cm]{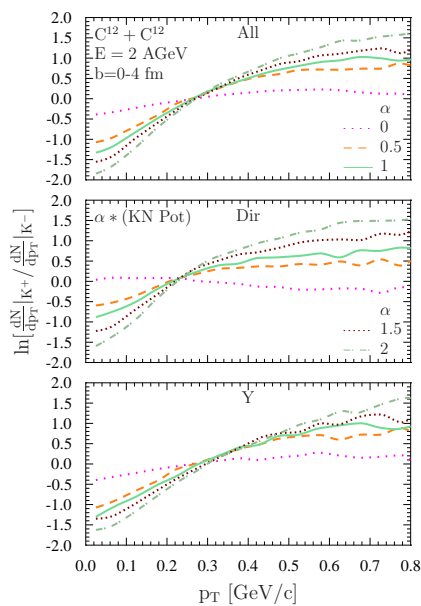}
\caption{\label{fig1} Logarithmic ratio of $p_{T}$ spectra of $K^+$ and $K^-$ for different strengths of potential. Various lines are explained in the figure.}
\label{fig2}
\end{figure}
In order to study this observable and in order to make sure that it does not depend on other input quantities 
we have separated the $K^-$ into 2 classes (by tracing back $K^-$ to its corresponding anti strange partner $K^+$).\\ (a) $K^-$ coming directly from  reactions like $BB \to BB K^+ K^-$ called direct contribution  and abbreviated in the figures (Dir)\\
(b) $K^-$ coming from $\pi Y$ or $BY\to K^-$ abbreviated in the figures by Y.\\
For the present study we use IQMD model the details of which are described
in ref. \cite{hartphysrep}. 
\section*{Results and Discussion} In fig. \ref{fig2} we display the logarithm of the ratio of the $p_{T}$ spectra of $K^+$ and $K^-$. Top, middle, and bottom panels show  this ratio for all $K^-$, for the directly produced $K^-$ and for those produced in secondary collisions, respectively. Different lines are for different strengths of potential which we vary by multiplying the K potential by a constant factor x. The total yields depend on the choice of x. 
The ratio is nearly constant without KN potential (x=0, dotted magenta line). When we switch on the potential the slope of the ratio changes very sharply in the low momentum region and decreases with increasing strength of the potential,
 whereas it remains nearly constant in the high momentum region. Comparing top and middle panel, we see that 
the influence of those $K^-$ (which come from secondary collisions) on the spectral form at small $p_t$ is not essential.
This means that this ratio is almost exclusively sensitive to the potential and does not depend on the little or unknown cross sections.
\begin{figure}[!t] \centering
\vskip 0.5cm
\includegraphics[angle=0,width=6cm]{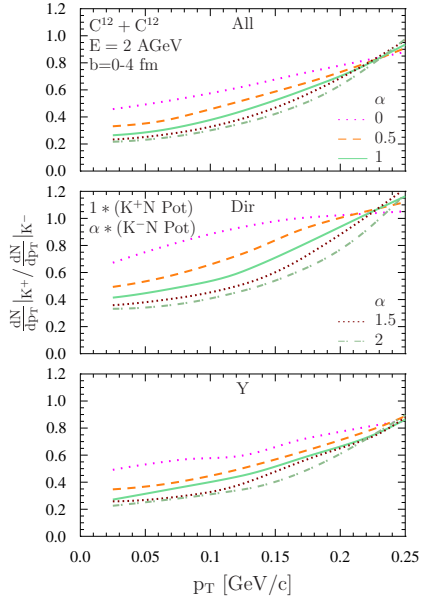}
\caption{\label{fig1} Same as fig. \ref{fig2} but only $K^-$N potential is varied for a fixed $K^+$N potential. }
\label{fig3}
 \vskip -0.3cm
\end{figure}
\begin{figure}[!t] \centering
\vskip 0.5cm
\includegraphics[angle=0,width=6cm]{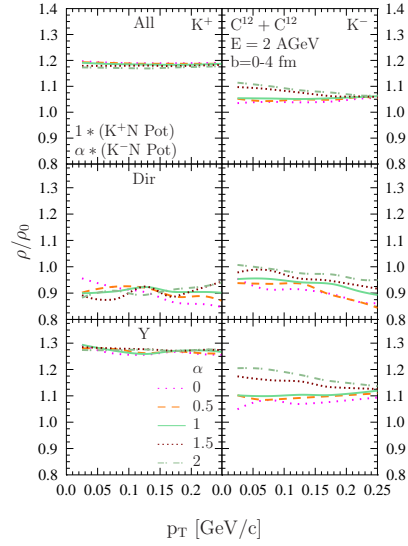}
\caption{\label{fig1} Density at which the finally observed kaons are produced. The various panels are explained in the text.}
\label{fig4}
 \vskip -0.3cm
\end{figure}
Fig. \ref{fig3}  presents as well the ratio of the $K^+$ and $K^-$  spectra but this time $K^+$N potential is taken as given
by the theoretical predictions (x =1) whereas for the $K^-$ we vary the potential assuming that the $K^+$ potentials
can be determined by other means. This time we have chosen a linear scale.
We observe, as expected,  that the dependence of the slope on the $K^-$ potential becomes weaker as compared to a variation of both potentials but still varies by a factor of two and is hence a measurable quantity. 
This ratio depends on the  $K^-$N potential only and presents therefore the possibility to measure directly the strength  of the $K^-$N potential. It is therefore the desired 'smoking gun' signal to determine experimentally the $K^-$ potentials in matter at finite densities.  
It is interesting to see at which density the kaons are produced which are finally seen in the detector. This is displayed in
fig. \ref{fig4}. On the left (right) hand side we display as a function of $p_T$ the average density  at which the $K^+$ ($K^-$) are produced which are finally seen in the detectors. The top panel shows the density for all events in which a $K^+$  and a $K^-$ is produced, the middle part that for those events in which the $K^+$  and a $K^-$ are produced simultaneously and the bottom part for those events in which the $K^-$ is produced in a secondary collision.
Independent of the potential the kaons are produced at densities around normal nuclear matter density. The density 
for the directly produced kaons  is slightly lower than that of the other events because the higher the density the higher is also the probability that the $K^-$ is reabsorbed in a $\Lambda$.

\section*{Acknowledgments}
This work has been supported by a grant from Indo-French Centre for the Promotion of Advanced Research (IFCPAR) under project no 4104-1.


\end{document}